\documentclass[pre,twocolumn,amsmath,aps]{revtex4}
\usepackage{graphicx}
\usepackage{dcolumn}
\usepackage{amssymb}
\usepackage{bm}
\usepackage{color}
\definecolor{white}{rgb}{1,1,1}

\begin{document}                         

\title{Growing networks under geographical constraints}
\author{R. Xulvi-Brunet$^{1,2}$ and I.M. Sokolov$^2$}
\affiliation{$^1$School of Mathematics and Statistics, University of Sydney,
             Sydney, NSW 2006, Australia\\
             $^2$Institut f\"{u}r Physik, Humboldt Universit\"{a}t zu 
             Berlin, Newtonstra\ss e 15, D-12489 Berlin, Germany}
\date{today}
% For large equations: \begin{widetext}  .... \end{widetext} 

\begin{abstract}
Inspired by the structure of technological networks, we discuss network 
evolution mechanisms which give rise to topological properties found in 
real spatial networks. Thus, the peculiar structure of transport and 
distribution networks is fundamentally determined by two factors. These 
are the dependence of the spatial interaction range of vertices on the 
vertex attractiveness (or importance within the network) and on the 
inhomogeneous distribution of vertices in space. We propose and analyse 
numerically a simple model based on these generating mechanisms which 
seems, for instance, to be able to reproduce known structural features of 
the Internet.
\vspace{0.4cm}
\vspace{0.7cm}
\hfil\break
\end{abstract}
\maketitle

Technological networks, such as transportation or communication networks, 
are man-made networks designed for transport of resources between sites 
distributed over a certain geographical area \cite{SIAM}. Depending on the 
type of network, the resources can be information, wares, electricity, or 
persons, and the geographical area range from a small region to the whole 
world. Examples of technological networks are, among others, the Internet 
\cite{Vazquez1,Yook1}, the airline \cite{Amaral1,Guimera1,Gastner1} and 
railway \cite{Latora1,Sen1} networks, and the electric power grid 
\cite{Amaral1,Watts1}. The Internet  consists of a set of routers linked 
by optical fibre or other type of physical connection, and it turned into 
an indispensable tool to get information from and about whatever part of 
the world. The airline network, which principal function is to transport 
persons and wares, has all the airports of the world as vertices of the 
network, and the corresponding nonstop scheduled flights connecting the 
airports as its edges. Electric power grids, on the other hand, are sets 
of generators, transformers, or substations connected by high-voltage 
transmission lines. 

The most prominent feature of these technological networks is that they are 
embbeded in a real physical space, with vertices having well-defined 
positions. This is not the case of other types of networks, such as citation 
or biochemical networks, in which the position of vertices has no physical 
meaning (see \cite{SIAM,Albert1,Dorogovtsev1,Boccaletti1} for general 
reviews). In many communication and transportation networks, the cost of 
establishing long-range connections between distant spots is usually higher 
than the cost of establishing short-range connections. This is clear for 
networks such as the Internet or the railway network, where establishing a 
long-range conection is obviously expensive because long channels need a 
larger infrastructure. For electric power grids, the connection cost between 
farther spots is even higher, given that in long high-voltage lines
a large amount of energy is lost during the transmission. 

This dependence of the connection cost on the distance is one of the most 
prominent mechanisms governing the evolution of technological networks, and 
it is determinant for understanding their structure. As a consequence of that, 
for example, not all connections between vertices are equally probable; 
neighbouring vertices tend to connect to each other with larger probability 
than distant ones. This, in turn, is the origin of some of their more 
characteristic properties.

The most important quantities designed for capturing network's structure 
are the degree distribution, the average distance between the vertices, and 
the mean and local clustering coefficients. The degree distribution $P(k)$ 
gives the probability that a randomly selected vertex of the network has 
degree $k$, i. e., that it is connected to $k$ other different vertices. 
Most technological networks exhibit degree distributions that decay as a 
power-law $P(k) \sim k^{-\gamma}$, i.e., they exhibit a scale-free character. 
However, power grids or railway networks, in which long-range connections 
practically do not exist, typically show exponential degree distributions 
\cite{Watts1,Sen1}. The average path length $l$ is defined as the mean 
distance between each two vertices in the network, where the distance between
any two vertices is defined as the number of edges along the shortest 
path connecting them. Finally, the clustering coefficients measure the local
tendency of vertices to form highly connected clusters. The clustering 
coefficient of a vertex is defined as the ratio between the number of 
connections existing among its nearest topological neighbours -the vertices 
which are connected through an edge with it- and the maximal number of edges 
which can exist among them. The mean and the local clustering coefficients, 
$C$ and $C(k)$, are the averages of the clustering coefficients over all 
vertices of the network, and over all vertices of degree $k$, respectively. 

Large mean clustering coefficients and average path lengths of many 
technological networks can be understood taking into account their growth 
mechanisms. Thus, the fact that vertices tend to link to their ``physical'' 
neighbours yields a large probability that, given a vertex in the network, 
its ``topological'' neighbours are also connected between themselves. That 
typically gives rise to large values of mean clustering coefficient. On the 
other hand, since distant vertices tend to be poorly connected between 
themselves, shortest paths connecting farther nodes are usually long, and 
pass through many vertices in between. Statistical measures of the lengths 
of connections confirm that the large part of edges in most transportation 
and distribution networks are short-range connections \cite{Gastner1}.

The distance between the vertices is an important parameter in the modelling 
of transportation networks, but the ``cost-distance'' dependence is not the 
only mechanism responsible for their structure. In general, the structure 
of technological networks is both a function of what is geographically 
feasible and what is technologically desirable. For reasons of efficiency, 
some long-range connections are typically always present, in spite of their 
high cost: in many cases, connecting two distant vertices trough a long 
necklace of neighbouring vertices slows down the global transport in the 
network and makes it inefficient. Long-range connections are observed both 
in the Internet and the Airline network \cite{Gastner1}. Additionally, when 
long-range connections exist, they usually link the highly connected vertices 
of the network \cite{Yook1,Gastner1}. That is not surprising. If a 
telecommunication company or an airline decides to make a big investment in
creating a long-range transport channel, it typically wants to link sites 
which are somehow important and well-connected (depending on the type of 
network, they can be technological, touristic or commercial spots), so that 
the amount of information or wares which will be exchanged between them 
compensates the expense.
 
Networks embedded in a metric space with distance-dependent connection
probabilities are called spatial or geographical networks \cite{Boccaletti1}. 
In the past few years several models have been proposed in order to study 
their structural properties 
\cite{Yook1,Xulvi-Brunet1,Manna1,Jost1,Rozenfeld1,Warren1,Dall1,Barthelemy1,
Sen2,Manna2,Hermman1,Avraham1,Kaiser1,Hayashi1,Hayashi2,Huang1,Andersson1,
Yang1,Massen1,Andrade1,Barrat1,Masuda1,Crucitti1,Roudi1,Cardillo1,Gastner2}. 
Most of them combine the preferential attachment mechanism \cite{Barabasi1}, 
which is widely accepted as the probable explanation for power law degree 
distributions seen in many networks, and distance effects. The last typically 
lead to a deviation from the scale-free behaviour when the distance 
constraints are sufficiently strong. Almost all these studies have focused 
on the effects of geography on the degree distribution, ignoring other 
important characteristics. These are, however, of primary interest. Thus, 
in a exhaustive study of the Internet \cite{Vazquez1} Vazquez and 
collaborators found not only that it is a scale-free network, but also that 
the local clustering coefficient $C(k)$ and the nearest neighbours' average 
function $\overline{k}_{nn}$ fall as power-law functions with exponents 
$-0.75$ and $-0.5$, respectively. On the other hand, Gastner and Newman 
\cite{Gastner1} showed that strong geographical constraints tend to produce 
networks with an effective network dimension $d$ close to $d \simeq 2$. These
new quantities are essential when measuring such important features as
degree-degree correlations and the hierarchy \cite{Vazquez1} and planar 
\cite{Gastner1} characters of networks.  

In this paper we propose several network evolving mechanisms which are 
able to develop spatial networks that exhibit most of the features found 
in real technological systems, i.e. reproducing correct values for 
$C(k)$, $\overline{k}_{nn}$ and $d$. The basic properties of our generating 
principles are: 

i) The knowledge of any given vertex about the network is 
limited to a certain (Euclidean) neighbourhood of the vertex (a property 
of locality). Each vertex is ``aware'' of the characteristics of all 
vertices belonging to its neighbourhood, but not of the characteristics 
of the rest of the vertices of the network. 

ii) The range of this physical neighbourhood is governed by a cost function 
which establishes the importance of the geographical constraints. As the 
connection cost grows, the range of the neighbourhood decreases. 

iii) As usual, the network grows by adding vertices and edges. At each time 
step, new vertices are added and connected to the system; additionally, new 
edges may be set between vertices already existing in the network. 

iv) Preferential attachment condition. Vertices try to connect to vertices of 
large degree  - the more attractive ones - lying within their neighbourhood. 

Apart from these requirements (already considered e.g. in 
Ref. \cite{Barthelemy1}), we add two new ingredients: 

v) The interaction cost governing the range of each neighbourhood depends on 
the attractiveness of the vertex associated; the larger is the vertex 
attractiveness the larger its interaction range. 

vi) The probability that a new vertex appears in an isolated area, 
geographically far from the rest of the vertices, is smaller than the 
probability that the new vertex appears close to an already existing vertex. 
(Regarding to this last point, a similar idea has recently been proposed by 
Kaiser and Hilgetag \cite{Kaiser1}). In addition, vertices cannot appear too 
close to each other. 

The last two conditions are inspired by the properties of technological 
networks. While condition v) is obviously a consequence of the fact that 
the degree of important vertices in the network grow faster the 
``richer'' they already are (preferential attachment), there is also another
reason. It is also because they extend their ``tentacles'' more far away. 
Condition vi) mirrors the fact that 
vertices do not appear over a geographical area at random. Consider, for 
instance, the Internet and the electric power grid. When a power plant 
is constructed in a region too far from civilisation, the plant supplies 
electricity to the buildings close to it, but no high-voltage transmission 
lines link the station to the grid of the civilised world if the distance is 
large; the plant usually remains isolated. (In fact, electric power plants 
are not constructed far from civilisation. The inhabitants of isolated regions 
usually use small generators for personal use. The station is constructed 
only when civilisation comes to the region.) In the Internet, routers 
concentrate in towns, rather than in deserted areas. That is quite natural, 
people live and work in towns! Consequently, new Internet accesses tend to 
appear in towns, in the vicinity of other already existing accesses. 
Furthermore, the more industrialised is the town, the more rapidly the number 
of Internet accesses grows. On the other hand, we must also take into account, 
that vertices do not appear extremely close to each other. Thus, 
constructing two big power stations in close vicinity to one another 
(for example, one kilometre apart) is not reasonable; it is cheaper to 
build one bigger station which supplies electricity to the entire area. It 
also is not common for a family house to have two routers, since one router 
can supply Internet to all computers in the house. Therefore, we assume that 
there are certain areas, not too far but also not too close to the existing 
vertices, where new vertices will more probably arise than in others.  

\onecolumngrid

\begin{figure}[htb]
  \centering
  \includegraphics[width=15.7cm]{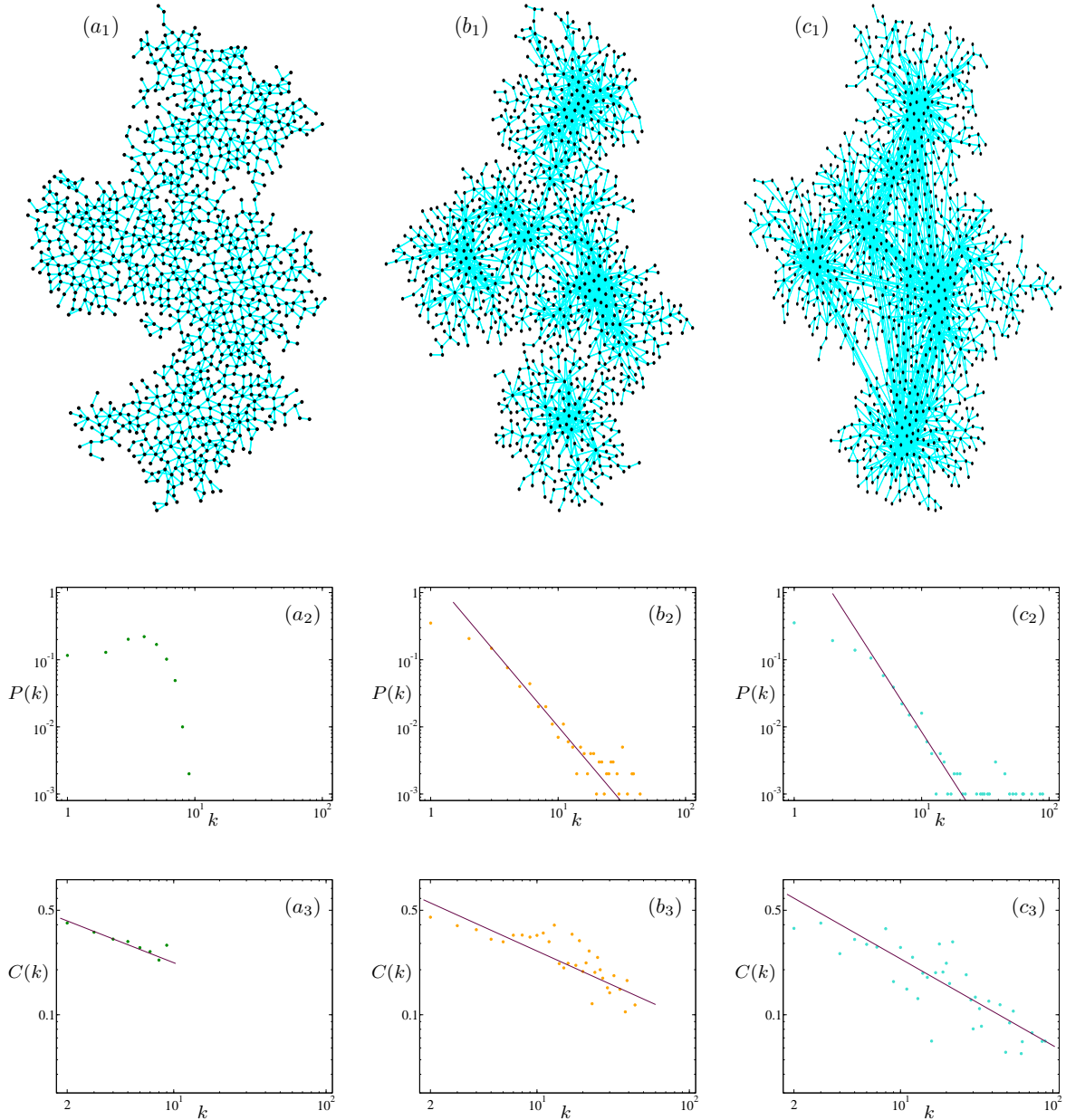}
    \caption{($a_1$), ($b_1$), and ($c_1$): graphical representations of the 
         models ($a$), ($b$), and ($c$), respectively. The three networks have
         $1000$ vertices; their size is $L=1881$ ($a_1$), $L=1982$ ($a_2$), 
         and $L=1982$ ($a_3$). Note that all edges of network ($a_1$) are 
         short-range connections, while in network ($c_1$) edges connecting 
         distant vertices do exist. ($a_2$), ($b_2$), and ($c_2$): Degree 
         distribution of the networks represented in ($a_1$), ($a_2$), and 
         ($a_3$), respectively. Models ($b$) and ($c$) are scale-free. The 
         slope of the straight lines are -2.24 
         ($b_2$) and -2.95 ($c_2$). ($a_3$), ($b_3$), and ($c_3$): Local
         clustering coefficient $C(k)$ of the networks represented in ($a_1$),
         ($a_2$), and ($a_3$), respectively. The behaviour of $C(k)$ follows
         power laws for three models. The slope of the straight lines 
         are -0.40 ($a_3$), -0.46 ($b_3$) and -0.58 ($b_3$). Notice the 
         double logarithmic scales in all graphs.}
    \label{fig1}
\end{figure}

\twocolumngrid

These basic ideas can be implemented in very different ways giving rise to 
different growth models. Consider, for instance, the preferential attachment 
prescription. One has to decide whether the attachment probability depends 
linearly on the degree, as in the Barab\'asi-Albert construction 
\cite{Barabasi1}, or whether it must follow another law. One also has to 
decide about the interaction's dependence on distance and on vertex 
attractiveness. Moreover, the probability of the appearance of a new vertex 
at position $\mathbf{r}$ can be a complex function involving the positions 
of all already existing vertices in the network, or depending only on the 
position of the vertices closer to the point. 

The appropriate implementation of a determined pattern depends obviously 
on the particular geographical system that we attempt to model. Numerical 
simulations show that different (but similarly oriented) prescriptions 
produce models showing qualitatively the same behavior. This fact supports 
the general value of the principles proposed. Since we are not yet intending 
to studying any particular real-world network, but are interested only in 
capturing some general features of spatial systems, we will adopt here the 
simplest realization of the guiding prescriptions.    

We start from a preselected area in a two-dimensional Euclidean plane. In 
this area, we place at random $m_o$ vertices, so that the distance between 
any two of these initial vertices is larger than a given $r_{min}$, the 
minimum distance that will separate vertices in the network. Since the 
$m_o$ vertices are placed at random, the order of magnitud of separation 
between them depends on the size of the preselected area. Now we let our 
network grow around these initial vertices. At each time step a new vertex 
with $m_1$ proper links is added to the network and connected to $m_1$ 
vertices already present. Additionally, once the new vertex is attached, 
$m_2$ new edges are distributed among all the vertices of the network. In 
both cases, vertices and edges are added to the network only if the 
geographical constraints allow for the addition. 

Our geographic constraints are defined by the two characteristic distances 
$r_{min}$ and $r_{max}$, which define a ring area around a point.  
At each step we choose at random a vertex of the preexisting network. 
From this point, using polar coordinates, we put a new vertex at a position 
given by a radius $r$ and an angle $\phi$ picked up at random from 
homogeneous distributions $r_{min} \le r \le r_{max}$ and 
$\phi \in \left] 0,2 \Pi \right]$. If this new vertex happens to be at a 
distance smaller than $r_{min}$ from some preexisting one, the selection is 
rejected and a different old vertex is chosen. Note that this prescription 
does not give an homogeneous distribution in space when $r_{max} \to \infty$ 
and $r_{min} \to 0$, but essencially means that smaller distancies to the 
choosen vertex are preferred.  

In order to connect the new vertex to the system, we consider the nodes 
of the network within the circle of radius $r_{max}$ from the 
newly introduced one $n$. This circular area around the new node is considered
to be its physical neighborhood. If the number of vertices in the neighborhood
is smaller than $m_1$, the newly introduced vertex is connected to all of them;
if their numer exceeds $m_1$ than it is connected to exactly the $m_1$ ones 
with higher degree. Note that the fact that the range of the neighbourhood is 
precisely $r_{max}$ ensures that a new node is connected to at least one old 
one. 

The second process, consisting of the addition of new edges between vertices, 
works in a similar way. We randomly choose a vertex $v$ of the network, and 
then, from the vertices that belong to its physical neighbourhood but are not 
yet connected to it, we choose the $m_2$ vertices having larger degree and 
connect $v$ to them. Here, however, the interaction range $r_v$ of the 
neighbourhood of $v$ is governed by the function
\begin{equation}
       r_v = r_{max}+ \beta {k_v}^\gamma \ ,       \label{rang}
\end{equation}
where $k_v$ is the degree of vertex $v$, and $\beta$ and $\gamma$ are 
non-negative tuning parameters whose function is basically to define the 
area which a vertex ``sees'' depending on its importance (degree) within 
the network. In case that the number of vertices that belong to the
neighbourhood of $v$, and that are not yet connected to it, is $q<m_2$, then
only $q$ edges are added to the network. Note that the effects of geography 
disappear when $r_{max} \to \infty$. 

Let us comment on some aspects of the model. First, vertices are not 
distributed at random over the area of study, but their distribution depends 
on the ``history'' of the network. New vertices appear in the vicinity of the 
vertices already present in the network. Second, the interaction range of a 
vertex is a function of its attractiveness, or importance in the network. If 
a vertex increases its importance in the network, then its interaction range 
grows. (Here, we do not take into account the fact that old vertices can 
remain obsolete with time.) Third, we impose the following preferential 
attachment condition: connect to the more attractive vertex of the network 
that you can ``see''. 

Extensive numerical simulations confirm that this simple model is able to
reproduce many of the properties of spatial networks. In the present study we 
restrict ourselves to two sets of values for the parameters of the model. The 
first one, which includes three different spatial cases, illustrates the 
impact of the cost-distance dichotomy on network structure. We consider the
following values: $m_1=1$, $m_2=1$, $r_{min}= 500$ m.u., and $r_{max}= 1000$ 
m.u. (where m.u. stands for an arbitrary ``metric unit''). The fact that we 
impose $r_{max}=2 \cdot r_{min}$, i. e., that $r_{max}$ is only twice 
$r_{min}$, indicates that in this case we deal with networks for which 
the cost of new vertices establishing long-range connections is very high. 
Additionally, we choose $m_0=7$, and a radius of $14000$ m.u. for our 
initially preselected disc-area, within which the $m_0=7$ initial vertices 
may be placed at random. The three cases we distinguish are: Case ($a$), 
$\beta=1$ and $\gamma=1.4$, corresponding to a spatial network in which the 
geographical constraints are extremely important (in this case long-range 
connections are practically inexistent); Case ($b$), $\beta = 1.5$ and 
$\gamma = 2.3$, an intermediate case; Case ($c$), $\beta=2$ and $\gamma =4$,
for which vertices of high degree are allowed to establish 
long-range connections. The selected values of parameters are certainly  
arbitrary and are adopted in order to illustrate the effects of the 
distance-cost-dependence. In effect, for $\beta = 0$, the resulting network 
is practically a tree, since no edges can be placed between old vertices, 
while for very large $\beta$ and $\gamma$ a ``winner-takes-all'' phenomenon 
emerges, in which almost all vertices are connected to one super-hub with 
an enormous degree.

Figure $\ref{fig1}$ compares the results of simulations corresponding to 
these three cases. To be able to draw the resulting networks, we 
consider small graphs with only $1000$ vertices (numerical simulations 
indicate that the structure does not significantly change as the order of the 
networks grows). Panels $a_1$, $b_1$, and $c_1$ show the effects of the 
selective growth of the interaction range with the degree of vertices: For 
systems where long-range connections are highly expensive (model $a$), 
even the most important vertices of the network are connected only to a few 
close neighbours. As the cost of establishing long-range interactions 
decreases, connections between distant vertices in the network begin to 
appear, in particular, between high degree vertices (models $b$ and $c$). 

\begin{figure}[tb]
  \centering
  \includegraphics[width=8.5cm]{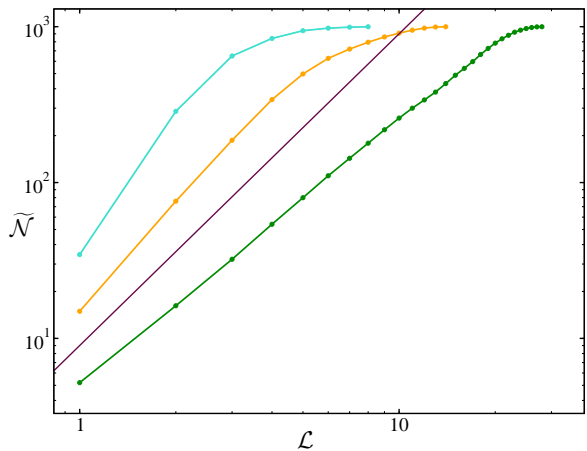}
    \caption{$\overline{\mathcal{N}}(\mathcal{L})$ as a function of the
         $\mathcal{L}$. From bottom to top, models $a$, $b$, and $c$. The 
         black straight line correspond to a network of dimension $2$. The 
         results are used to estimate the dimension of the three spatial 
         networks considered (see text for more details). Notice the double 
         logarithmic scales of the picture.}
    \label{fig2}
\end{figure}

The degree distribution evidently changes as the geographical constraints 
are gradually loosened. Thus, model $a$ shows a degree distribution which 
decays approximately exponentially (panel $a_2$); no vertices 
of high degree can be found. The degree distributions of models $b$ and $c$, 
however, exhibit well-defined power law tails (in spite of the small order 
of the networks considered): $P(k) \sim k^{-2.25}$ (panel $b_2$) and 
$P(k) \sim k^{-2.95}$ (panel $c_2$). The corresponding local clustering 
coefficients also show a behaviour very close to that found in 
real networks (see panels $a_3$, $b_3$, and $c_3$). All three models exhibit
power law behaviours for $C(k)$: $C(k) \sim k^{-0.40}$ (panel $a_3$), $C(k) 
\sim k^{-0.46}$ (panel $b_3$), and $C(k) \sim k^{-0.58}$ (panel $c_3$). In
addition, the mean clustering coefficient $C$ of these spatial models is 
always quite large, about $C \simeq 0.33$ for all three of them. The number of 
triangles (cycles of length three) in the network is, however, $668$ 
(model $a$), $1593$ (model $b$), and $1341$ (model $c$). On the other hand, 
the average path length decreases as the amount of long-range connections 
grows from $l=20.18$ (model $a$)to $l=8.83$ (model $b$) and $l=5.01$ 
(model $c$). This result is quite natural, and shows the transition from a 
quasi-planar graph with a structure quite similar to a lattice (model $a$) 
to a typical complex network structure found in most geographical networks 
(models $b$ and $c$).       

\begin{figure}[tb]
  \centering
  \includegraphics[width=8.5cm]{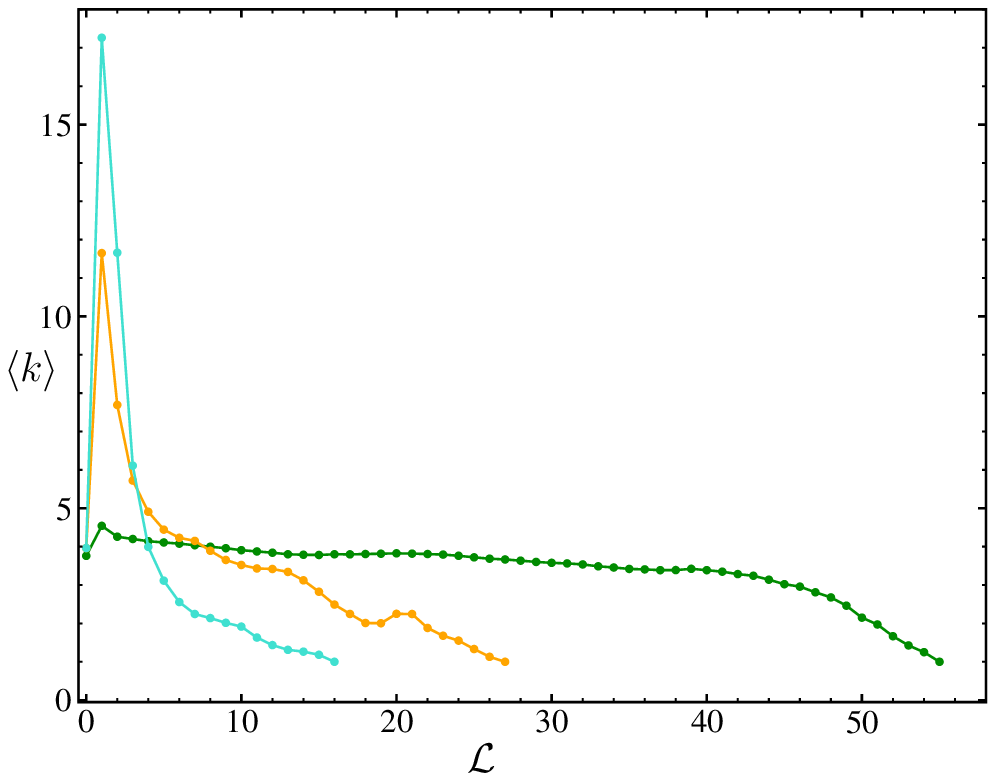}
    \caption{Layer average degree, $\langle k \rangle_{\mathcal{L}}$, as a
         function of shell number $\mathcal{L}$.
         The curve with the highest peak corresponds to model $c$, the 
         intermediate one 
         and the flat one to models $b$, and $a$, respectively.}
    \label{fig3}
\end{figure}

Making use of the effective dimension $d$, Gastner and Newman showed that, 
networks where geographical effects are extreme are 
essentially planar graphs (i.e., they can be drawn on a map without any 
edges crossing). The effective dimension can be defined as 
$d = \lim_{\mathcal{L} \to \infty} \log{\overline{\mathcal{N}}
(\mathcal{L})} / \log{\mathcal{L}}$, where 
$\overline{\mathcal{N}}(\mathcal{L})$ 
is the average number of vertices which can be found within a distance of 
$\mathcal{L}$ steps or less from a vertex. In finite networks no limit 
$\mathcal{L} \to \infty$ can be taken, but good results for $d$ can be 
achieved by plotting $\log{\overline{\mathcal{N}} (\mathcal{L})}$ against 
$\log{\mathcal{L}}$ for the central vertices of the network and
measuring the slope of the resulting line (of course, far away of the 
saturation region corresponding to exhausting of the network). Central 
vertices are those vertices of the network that have minimum eccentricity, 
being defined as the maximum distance from the vertex to any other vertex in 
the network. (Note that central vertices are sometimes defined as the
vertices having larger ``betweenness centrality'', as in \cite{Guimera1}. 
We use, however, the classical definition from the graph theory.) 

Figure $\ref{fig2}$ shows on double logarithmic scales how 
$\overline{\mathcal{N}}(\mathcal{L})$ behaves as a function of 
$\mathcal{L}$. From bottom to top the curves correspond to model ($a$), 
the critical behaviour $\overline{\mathcal{N}}(\mathcal{L}) \sim 
{\mathcal{L}}^2$, model ($b$), and model 
($c$). We see that the effective dimension of model $a$ is certainly 
smaller than two, which is not a surprize provided that model $a$ 
creates practically a planar graph. The dimensions of models $b$ and $c$ 
-which are obviously not planar graphs- are larger than two. The difference 
between the three models at $\overline{\mathcal{N}}(\mathcal{L}=1)$ is due 
to the fact that central vertices are usually the more connected ones of the 
network.

\begin{figure}[tb]
  \centering
  \includegraphics[width=8.5cm]{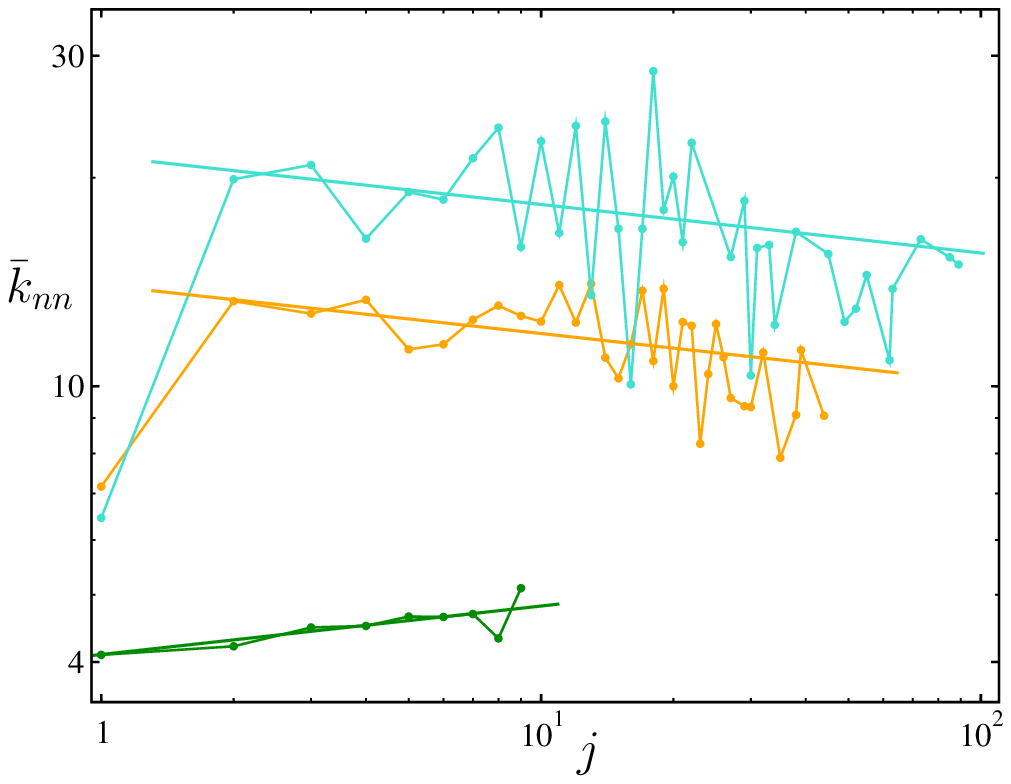}
    \caption{Analysis of the degree-degree correlations of the three models 
             considered. Nearest neighbours' average function 
             $\overline{k}_{nn}(j)$ against $j$. From bottom to top, models 
             $a$, $b$, and $c$. Note that models $b$ and $c$ exhibit 
             dissortative mixing, while model $a$ is slightly assortative.}
    \label{fig4}
\end{figure}

Tomographic studies reveal interesting details too. Tomography deals 
with the study of the structure of layers which surround a given vertex (the 
root) in the network \cite{Kalisky1,Xulvi-Brunet2,Volz1}. The principal 
motivation for examining the tomography of a network results from its 
importance for understanding the spreading phenomena taking place in 
networks. We concentrate here on the layer average degree 
$\langle k \rangle_{\mathcal{L}}= \sum_k k P_{\mathcal{L}}(k)$, 
where $P_{\mathcal{L}}$, the degree distribution in shell $\mathcal{L}$, is 
defined as $P_{\mathcal{L}}(k)=(\sum_r N_{\mathcal{L},r}(k))/(\sum_{k,r} 
N_{\mathcal{L},r}(k))$, with $N_{\mathcal{L},r}(k)$ being the number of 
vertices of degree $k$ in layer $\mathcal{L}$ for root $r$. The study of
$\langle k \rangle_{\mathcal{L}}$ for the three networks considered shows a 
peak whose height decreases as the cost of establishing long-range connections
grows (fig. $\ref{fig3}$). The results indicate that the mean degree 
$\langle k \rangle_{\mathcal{L}=1}$ increases rapidly as the number of 
long-range connections and hubs in the network grows. On the other hand, the 
average shell degree decreases rapidly for more distant layers, 
$\mathcal{L}>1$. This interesting result shows that vertices with large 
degrees are rapidly exhausted in this type of networks, which has especial 
importance when dealing with spreading phenomena, like spreading of 
information or infections. Note that this result has important effects on 
epidemiological properties: vertices of large degree are rapidly affected 
by the spreading of an infection. On the other hand, in a network like that 
of model ($a$) the propagation of any spreading agent will be similar to the 
propagation on a lattice: the spreading agent will primarily reach the 
nearest physical neighbours. 

In figure $\ref{fig4}$ we show the correlation properties of these three 
models. Degree-degree correlations are determined by the probability 
function $\mathcal{E}_{ij}$, which gives the probability that a randomly
selected edge connects one vertex of degree $i$ to another of degree $j$. 
Thus, a network is said to be degree-degree uncorrelated if 
$\mathcal{E}_{ij}=(2-\delta_{ij})iP(i)jP(j)/\langle i \rangle^2$, which 
only means that the probability that an edge connects 
to a vertex of a certain degree $k$ is independent from whatever vertex is 
attached to the other end if the edge. Otherwise, the network is said to 
be degree-degree correlated. Most real networks are correlated, and usually
exhibit either ``assortative'' or ``dissortative'' mixing \cite{Newman1}. 
Assortativity means that high-degree vertices attach preferably 
to other highly connected vertices, i. e., with a larger probability than 
in uncorrelated networks; on the other hand, dissortativity stands for 
when high-degree vertices tend to connect to low-degree vertices, and vice 
versa. Thus, a very useful quantity for measuring the correlation's degree 
of a network is the nearest neighbours' average function 
$\overline{k}_{nn}(j)$, which expressed in terms of $\mathcal{E}_{ij}$, 
can be written as $\overline{k}_{nn}(j)=(\sum_i 
i(1+\delta_{ij})\mathcal{E}_{ij})/(\sum_i (1+\delta_{ij})\mathcal{E}_{ij})$. 
It takes the constant value $\overline{k}_{nn}(j)=\langle j^2 \rangle / 
\langle j \rangle$ if no type of degree-degree correlation exist, while it 
is a decreasing (increasing) function if dissortative (assortative) 
mixing is present. In the picture we plot $\overline{k}_{nn}(j)$ as 
a function of $j$. The lowest curve, corresponding to model ($a$), 
shows then that the network is slightly assortative. This feature of model 
($a$) is due to the fact that the areas containing a large density of 
vertices usually contain a large density of edges (see Figure 
$\ref{fig1}$, $a_1$), 
corresponding probably to important areas of the space; and vice 
versa, the areas containing a small density of vertices also contain a small 
density of edges. On the other hand, models ($b$) and ($c$), in which the 
geographical constraints are not so strong, present dissortative mixing. 
Interestingly, for both models $\overline{k}_{nn}(j)$ falls with $j$ 
following power laws of the form $\overline{k}_{nn}(j) \sim j^{-\iota}$, 
just like happens in real networks. 

Model ($a$) reproduces quite well the properties of those systems where 
vertices and edges are embedded in the two-dimensional physical space, 
like, for example, electric power grids or road networks. However, none 
of the three models considered above is suitable for characterising 
world-scale systems such as the Internet or the network of airline routes. 
The reason is that, in such large-scale systems, vertices are usually not 
uniformly distributed in the region under study (as occurs in our preceding 
models, see pictures \ref{fig1} $a_1$, $b_1$, and $c_1$), but they concentrate 
in a number of technological areas distributed over the world. Thus, a more 
realistic model for describing such systems must take into account that 
in large-scale geographical networks there usually exist many ``desert'' 
regions lying between the areas where vertices can be found in abundace. 
Such a pattern is easy to construct by varying the ratio between $r_{max}$ 
and $r_{min}$ in our model. This aspect is actually considered by our second 
selection of parameters. As we will see next, inhomogeneous distribution of 
vertices in space influences quantitatively the statistical properties of 
networks.

\onecolumngrid

\begin{figure}[!]
  \centering
  \includegraphics[width=15.7cm]{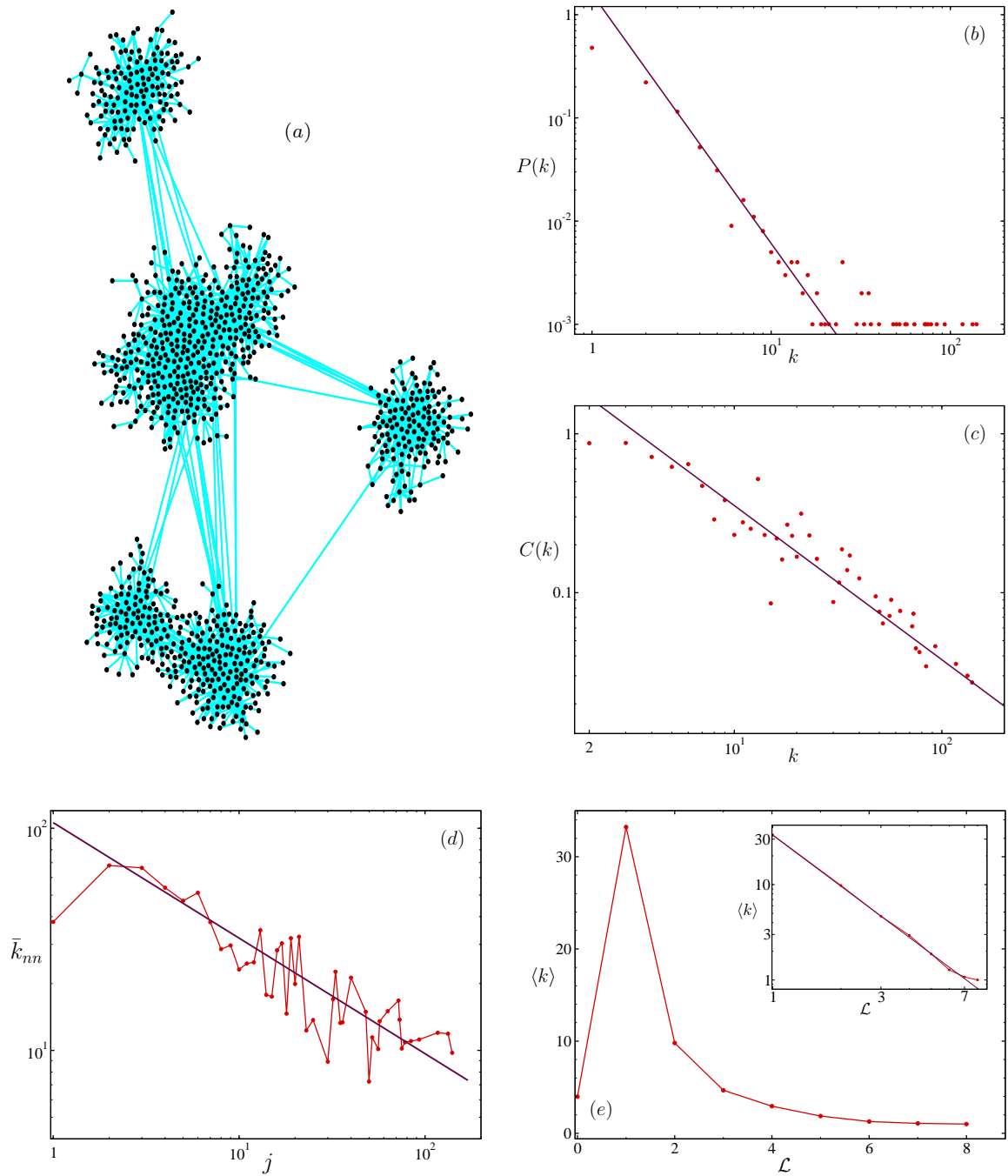}
    \caption{($a$): Graphical representation of a small network ($N=1000$ 
         and $L= 
         1985$) corresponding to model ($d$). Note that vertices concentrate 
         in certain areas of the space, and that the long-range connections
         of the network link end-vertices of large degree. ($b$): Degree
         distribution of model ($d$). ($c$): Degree-dependent clustering
         coefficient $C(k)$ of the model. ($d$): nearest neighbours' average 
         function $\overline{k}_{nn}$ as function of the degree $j$. Note that
         $P(k)$, $C(k)$, and $\overline{k}_{nn}$ fall off as power law 
         functions. 
         ($e$): Average degree, $\langle k \rangle_{\mathcal{L}}$, as a
         function of shell number $\mathcal{L}$. From $\langle k 
         \rangle_{\mathcal{L}=1}$ the average degree decays interestingly 
         following a power law (inset of the picture).}
    \label{fig5}
\end{figure}

\twocolumngrid

\textcolor{white}{blanc blanc blanc blanc blanc blanc blanc blanc blanc}\\

Let us thus consider a larger ratio for $r_{max}/r_{min}$, for example, 
$r_{max}=5 \cdot r_{min}$. (Note that the change in 
the ratio $r_{max}/r_{min}$ modifies not only the distribution of vertices 
in space, but also makes cheaper the cost of establishing connections for 
new vertices, since the neighbourhood of a new vertex will, in comparison, 
contain more vertices.) The values of the parameters for this model 
(model $d$) are now the following: $m_0=7$, $m_1=1$, $m_2=1$, $r_{min}= 200$ 
m. u., $r_{max}= 1000$ m. u., $\beta=2$ and $\gamma=3$. As before, the initial 
area, where the $m_0=7$ vertices are randomly placed, has a radius of 
$14000$ m. u. and the order of the network is $N=1000$. Figure 
$\ref{fig5}$\ ($a$) shows this network in the two-dimensional Euclidean space: 
The model simulates perfectly the tendency of vertices to concentrate in 
different areas having a high density of vertices (as if these areas were 
urban centres, i.e., cities, or city agglomerations), which are linked 
through long-range connections which join vertices of large degree, usually 
belonging to different geographical communities. 

The properties of this last construction are especially interesting, since 
they reproduce many structural properties found in the Internet. Thus, the 
degree distribution of the model follows a power law 
$P(k) \sim k^{-2.42}$ (panel $b$ of figure
$\ref{fig5}$). The mean clustering coefficient is large. For our small 
network $C \simeq 0.7$. The degree-dependent clustering 
coefficient $C(k)$ decays as a power law function $C(k) \sim k^{-0.97}$ 
(panel $c$ of the Figure). The decay of $C(k)$ gets however
slower for networks of larger size $N$. The average path length is very small, 
$l=3.74$ for the parameters as in Figure $\ref{fig5}$, and numerical 
simulations with larger networks indicate a small-world behaviour. In 
addition, the network shows dissortative mixing: the nearest neighbours' 
average function $\overline{k}_{nn}$ 
decreases as $\overline{k}_{nn}(j) \sim j^{-0.52}$ (see Fig. \ref{fig5} $d$). 
The coincidence of these properties with the ones of the Internet 
is astonishing: Let us now remind that i) the degree distribution of the 
Internet follows a power law with exponent $\gamma \simeq -2.5$, ii) the 
local clustering function $C(k)$ behaves as $C(k) \sim k^{-0.75}$, and iii) 
$\overline{k}_{nn}$ decreases with $j$ following the function 
$\overline{k}_{nn}(j) \sim j^{-0.5}$ \cite{Vazquez1}.  Finally, in Figure 
\ref{fig5} we plot 
($e$) the average degree $\langle k \rangle_{\mathcal{L}}$ as a function of
shell number $\mathcal{L}$, corresponding to the study of tomography.
We see again that hubs are found only a few steps away from any vertex, and 
interestingly, that $\langle k \rangle_{\mathcal{L}}$ drops as a perfect 
power law from $\mathcal{L}=1$ on (see inset of the Figure; note the
double logarithmic scales). 

To conclude, we introduce several network-generating mechanisms taking into 
account the constraints that geography impose on the evolution of large-scale 
network systems in physical space. We suggest that two properties are 
determinant 
for the structure of such geographical networks: the fact that the 
spatial interaction range of vertices depends on the vertex attractiveness 
and the fact that that vertices are not randomly distributed in space. 
Simple implementations of these mechanisms show that
%, apart from efficiency or economico-political design reasons,% 
the essential difference between 
``strong geographical'' networks, such as electric power grids, and ``weak
geographical'' networks, such as the Internet or the airline network, 
could be the cost (economical or technological) of establishing long-range 
connections. On the other hand, inhomogeneous distribution of vertices in 
large-scale networks seems certainly to be a relevant generating element of 
their hierarchical character. In any case, the agreement of our results 
with the properties found in real networks suggest that the mechanisms 
proposed may play a key role in the evolution and structure of networks.

\end{document}